\documentstyle[12pt]{article}

\newcommand{\sect}[1]{\setcounter{equation}{0}\section{#1}}

\def\beq{\begin{equation}}
\def\eeq{\end{equation}}
\def\bea{\begin{eqnarray}}
\def\eea{\end{eqnarray}}
\def\bq{\begin{quote}}
\def\eq{\end{quote}}

\renewcommand{\a}{\alpha}
\renewcommand{\b}{\beta}

\renewcommand{\d}{\delta}
\newcommand{\D}{\Delta}

\newcommand{\e}{\epsilon}

\newcommand{\n}{\nu}

\newcommand{\r}{\rho}

\newcommand{\s}{\sigma}

\renewcommand{\o}{\omega}
\renewcommand{\O}{\Omega}

\newcommand{\half}{\frac{1}{2}}
\newcommand{\intsumk}{\sum_k}
\newcommand{\intsuml}{\sum_l}

\newcommand{\an}{a_{2n-1}^{(2)}}
\newcommand{\bn}{b_n^{(1)}}
\newcommand{\cn}{c_n^{(1)}}
\newcommand{\dn}{d_{n+1}^{(2)}}

\newcommand{\mt}{\check{m}}

\newcommand{\reff}[1]{(\ref{#1})}

\begin{document}

\newpage
\begin{titlepage}
\begin{flushright}
{KCL-TH-94-18}\\
{BRX-TH-362}\\
{hep-th/9411176}
\end{flushright}
\vspace{2cm}
\begin{center}
{\bf {\large TODA SOLITON MASS CORRECTIONS\\
AND THE PARTICLE--SOLITON DUALITY CONJECTURE}}\\
\vspace{1.5cm}
G.W. DELIUS\footnote{Supported by Habilitationsstipendium der
Deutschen Forschungsgesellschaft}
\footnote{On leave from Department of Physics, Bielefeld University,
Germany}\\
\vspace{2mm}
{\em Department of Mathematics}\\
{\em King's College London}\\
{\em Strand, London WC2R 2LS, UK}\\
{\small e-mail: delius@mth.kcl.ac.uk}\\
\vspace{5mm}
M.T. GRISARU\footnote{Work partially supported by the
National Science Foundation under
grant PHY-92-22318.} \\
\vspace{2mm}
{\em Brandeis University}\\
{\em Waltham, MA 02254, USA}\\
{\small e-mail: grisaru@binah.cc.brandeis.edu}\\

\vspace{1.6cm}
{\bf{ABSTRACT}}
\end{center}
\bq
We compute quantum corrections to soliton masses in affine Toda
theories with imaginary exponentials based on the nonsimply-laced
Lie algebras $c_n^{(1)}$.
We find that the soliton mass ratios renormalize  nontrivially,
in the same manner as those of the fundamental particles of the
theories with real exponentials based on the  nonsimply-laced
algebras $b_n^{(1)}$.
This gives evidence that the conjectured relation between solitons in one
Toda theory and fundamental particles in a dual Toda theory holds also at
the quantum level. This duality can be seen as a toy model for S-duality.
\eq

\vfill

\end{titlepage}

\sect{Introduction.}

For the past few years affine Toda theories have provided an interesting arena
for studying properties of two-dimensional integrable field theories. They
are theories of $r$  real bosonic fields $\vec{\phi}$ coupled via exponential
interactions, with lagrangians of the form
\beq\label{lag}
{\cal L}=-\half \vec{\phi} \cdot \Box \vec{\phi} - \frac{m^2}{\tilde{\b^2}}
 \sum_i n_i\,\left(e^{\tilde{\b} \vec{\a}_i \cdot\vec{\phi}}-1\right)
\eeq
where the $\vec{\a}_i$ are the simple roots of a rank $r$ affine Lie algebra
$\hat{g}$, the $n_i$ are Ka\v{c} labels, $m$ sets the mass scale and
$\tilde{\b}$ is a coupling constant. For real $\tilde{\b}=\b$ the lagrangians
describe systems of $r$ massive particles created by the fields $\vec{\phi}$.
For purely imaginary $\tilde{\b}=i\b$ the theories possess soliton solutions.

The knowledge of the Toda theories is most complete in the real $\tilde{\b}$
regime. Because the theories possess higher spin conserved charges
\cite{Wil81,Oli83,Oli85,Oli86},
even at the quantum level \cite{Del92a,Fei92,Nie94}, the S-matrices for the
scattering of
the elementary particles are factorizable and have been determined exactly
\cite{Bra90,Del92b}. One makes the remarkable observation \cite{Del92b} that
the S-matrix of the
Toda theory for one Ka\v{c}-Moody algebra $\hat{g}$ at real coupling $\b$ is
equal
to the S-matrix of the Toda theory for the dual Ka\v{c}-Moody algebra
$\check{\hat{g}}$ at coupling $4\pi/\b$. Thus there is a strong
coupling --
weak coupling duality.

Many elegant results have also been obtained about the classical soliton
solutions of the Toda theories with imaginary $\tilde{\b}$
\cite{Hol92}--\cite{Zhu93}.
%,Oli93a,Oli93b,Kne93,Fri94,Kne94,Ara93,Mac92,McG93,Und93,
In particular their masses and more recently \cite{Fre94} the values of the
higher spin
conserved charges have been calculated. Again one makes another
observation \cite{Oli93a} of a different duality: the classical
masses and charges of the solitons of the Toda theory based on an untwisted
Ka\v{c}-Moody algebra $g^{(1)}$ are proportional to the unrenormalized masses
and charges
of the fundamental particles of the Toda theory based on a dual untwisted
Ka\v{c}-Moody algebra $\check{g}^{(1)}$.

These dualities are in close analogy to the conjectured duality between
the magnetically charged monopoles and the electrically charged gauge
particles in Yang-Mills theory \cite{God77,Mon77}. In its strong form
this conjecture states that the magnetic monopoles of one theory can
be reinterpreted as the gauge particles of another Yang-Mills theory
with a dual gauge group and with the inverse coupling constant. This
conjecture has recently become important because it is a consequence
of S-duality in string theory \cite{Sen94,Vaf94}. The relation in affine
Toda theory between soliton and particle masses and the duality which relates
strong and weak coupling while interchanging an algebra and its dual algebra
\cite{Del92b} can
be seen as a simplified laboratory for these ideas.

The Olive-Montonen duality conjecture \cite{God77,Mon77} was based on the
classical mass-relation between the monopoles and the gauge particles. The
conjecture was then further pursued only in cases when supersymmetry forbade
any quantum corrections to the mass formulas \cite{Wit78,Osb79}, so that the
classical relation was guaranteed to persist. It was feared that any quantum
corrections would spoil the duality; however,  explicit calculations were too
difficult
to perform. In this paper we will perform calculations in the simpler
Toda theories and see that here the relation between soliton and
particle masses persists even in the presence
of non-trivial quantum corrections.

We will calculate the first quantum corrections to
the masses of the solitons of the $c_n^{(1)}$ Toda theory. We will find that
the mass ratios renormalize non-trivially. Indeed they renormalize in the same
way as the mass ratios
of the fundamental particles of the (dual) $b_n^{(1)}$ Toda theory found in
\cite{Del92b}. Thus the duality relation is preserved, at least at the first
order in quantum corrections.

In \cite{Hol93a} Hollowood has calculated the first quantum correction to the
masses of the solitons of  (self-dual) $a_n^{(1)}$ Toda theory. He finds that
in that case
the mass ratios do not receive quantum corrections. This phenomenon of
non-renormalization of mass-ratios holds also for the fundamental
particles of all Toda theories based on self-dual Ka\v{c}-Moody algebras.
Thus Hollowood's result is in agreement with the quantum duality
conjecture.

In the case of $c_2^{(1)}$ Toda theory quantum corrections to the soliton
masses have  been computed by Watts \cite{Wat94} using essentially the same
techniques.
Our results are in disagreement with his when specialized to the $c_2^{(1)} $
case.

We use techniques similar to those employed by Hollowood \cite{Hol93a}, as
borrowed from
Dashen et al. \cite{Das74}.
The classical soliton states for the nonsimply-laced case,
given by the Hirota solution, are
constructed by folding from the simply-laced ones. Quantum corrections are
obtained by looking at zero-point energies for oscillations around the
classical solutions.

Our paper is organized as follows: in the next section we describe the
classical soliton solutions for the $c_n^{(1)}$ theories.
In section 3 we discuss the spectrum of perturbations
around these classical solutions and compute the mass
corrections, the result being given in \reff{result} and \reff{ratios}.
Section 4 contains a discussion.

\sect{Classical solutions in the $c^{(1)}_n$ theories}

In this section, primarily to fix the notation,  we review the Hirota
construction of the classical solutions
of Toda theories. The solutions for the nonsimply-laced algebras can be
obtained by folding from those for the simply-laced algebras. In the present
case we have constructed the soliton solutions of the $c_n^{(1)}$ theories from
the
 solutions of the $a^{(1)}_{2n-1}$ theories described by
Hollowood \cite{Hol92}.
Other methods exist now for achieving the same goal \cite{Ara93}.

The classical field equations of the Toda theories are given by
\beq\label{e21}
\Box \vec{\phi} = - \frac{m^2}{i\b} \sum_{j=0}^rn_j \vec{\a}_j e^{i\b
\vec{\a}_j \cdot \vec{\phi}}
\eeq
The Hirota solution starts with the Ansatz
\beq\label{phitau}
\vec{\phi} = \frac{i}{\b} \sum_{j=0}^r \frac{2\vec{\a}_j}{\a_j^2} \ln \tau_j
\eeq
with the functions $\tau_j (x,t)$ to be determined.
Since the set of roots $\vec{\a}_j$, $j=0,1,\dots,r$ is overcomplete the
Ansatz has an
invariance which can be used to make a gauge choice
\beq\label{gauge}
D^2 \tau_0 = m^2 \frac{\a_0^2}{2} \left( \prod_{k=0}^r \tau_k^{-K_{k0}
+2\d_{k0}} -\tau_0^2 \right)
\eeq
where
\beq
D^2\tau \equiv \ddot{\tau}\tau - \dot{\tau}^2 - \tau'' \tau +\tau'^2.
\eeq
$K_{jk} \equiv 2\vec{\a}_j \cdot \vec{\a}_k/\a_j^2$
is the Cartan matrix, and dots and primes indicate time and space derivatives
respectively.
Substituting (2.2) and (2.3) into (2.1) gives the $\tau$ equations of motion
\beq\label{eom}
D^2 \tau_j= m^2 \frac{\a_j^2}{2} n_j \left( \prod_{k=0}^r \tau_k^{-K_{kj}
+2\d_{kj}} -\tau_j^2 \right) ~~~,~~~ j=1,\dots,r
\eeq
which therefore take the same form as the gauge condition in (2.3).

The $a_{2n-1}^{(1)}$ theory has rank $r=2n-1$, and roots of equal length
$\a_j^2=2$ for all $j=0,1,\dots,2n-1$. There exist $2n-1$ single soliton
solutions
of the equations given by \cite{Hol92}
\beq
\tau_j^{(a)}= 1+e^{\O_a}\o^{ja} ~~~,~~~ a=1,\dots,2n-1
\eeq
with
\beq\label{e27}
\O_a = \s_a (x-v_at) - \xi_a,\qquad,\qquad
\o=e^{\frac{\pi i}{n}}
\eeq
and
\beq\label{smass}
\s_a^2 (1-v_a^2) =4m^2 \sin^2\frac{a\pi}{2n} \equiv m_a^2
\eeq
$v_a$ is the velocity, $1/\sigma_a$ the width of the soliton and relation
(\ref{smass}) expresses the Lorentz contraction.
The real part of $\xi_a$ determines the center of mass position, while the
imaginary part (which falls into $2n-1$ equivalence
classes) determines the asymptotic values of the soliton solution,
and therefore its topological  charge \cite{Hol92,McG93,Und93}.

The Hirota Ansatz gives two-soliton solutions by a nonlinear superposition
\beq\label{twosolitons}
\tau_j^{(ab)}= 1+ e^{\O_a}\o^{ja} +e^{\O_b}\o^{jb}
+A_{ab}e^{\O_a+\O_b}\o^{j(a+b)}
\eeq
with the ``interaction'' function
\beq\label{intfunc}
A_{ab}= -\frac{(\s_a-\s_b)^2-(\s_av_a-\s_bv_b)^2-m^2_{a-b}}
{(\s_a+\s_b)^2-(\s_av_a+\s_bv_b)^2-m^2_{a+b}}
\eeq

The single soliton solutions for the $c_n^{(1)}$ theory are constructed from
these two-soliton solutions.
We denote the roots of $c_n^{(1)}$ by $\tilde{\vec{\a}}$:
\bea
\tilde{\vec{\a}_0} &=& \vec{\a}_0 ~~~,~~~ \tilde{\vec{\a}_n} = \vec{\a}_n
\nonumber\\
\tilde{\vec{\a}_j} &=& \frac{1}{2}(\vec{\a}_j +\vec{\a}_{2n-j})
\eea
It follows then from the general Ansatz (2.2) that for any Hirota solution
of the $a^{(1)}_{2n-1}$ theory with $\tau_j=\tau_{2n-j}$ we obtain a solution
of
the $c_n^{(1)}$ theory
\beq
\tilde{\vec{\phi}} = \sum_{j=0}^n \tilde{\vec{\a}}_j
\frac{2}{\tilde{\a}_j^2} \ln \tilde{\tau}_j
\eeq
by setting $\tilde{\tau}_j = \tau_j$, so that $\tilde{\vec{\phi}}=\vec{\phi}$.
Solutions $\vec{\phi}$ of the $a_{2n-1}^{(1)}$ theory satisfying
$\tau_j=\tau_{2n-j}$ are given by
the two-soliton solutions \reff{twosolitons}
with soliton $a$ and soliton $2n-a$ moving
together, i.e. with
\beq
b=2n-a ~~~,~~~ \O_b =\O_a
\eeq
Thus, the $n$ one-soliton solutions of the $c_n^{(1)}$ theory are given by
\beq
\tilde{\tau}_j^{(a)}= 1+e^{\O_a} \d_j^{(a)} +e^{2\O_a}A^{(a)}
{}~~~,~~~a=1,\dots,n
\eeq
with
\bea
\d_j^{(a)} &=& \o^{ja}+\o^{j(2n-a)} =2\cos\left(\frac{\pi}{n}aj\right)
\nonumber\\
A^{(a)} &=& A_{a,2n-a} = \cos^2 \left( \frac{\pi}{2n}a\right)
\eea
and $\O_a$ given by (2.7).
We note that for $a=n$ the solution for the $c_n^{(1)}$ theory is
identical to that for the $a_{2n-1}^{(1)}$ theory (up to a shift in $\xi$):
\beq
\tilde{\tau}_j^{(n)}=1+2e^{\O_n}(-1)^j
\eeq

The two-soliton solutions $\tilde{\tau}^{(ab)}$ of the $c_n^{(1)}$ theory are
similarly obtained from the four-soliton solutions of the $a_{2n-1}^{(1)}$
theory
\bea\label{ctwosol}
\tilde{\tau}_j^{(ab)} &=& 1+e^{\O_1}+e^{\O_2}+e^{\O_3}+e^{\O_4} \nonumber\\
&& +e^{\O_1+\O_2} A_{12}+ e^{\O_1+\O_3} A_{13} + \cdots \nonumber\\
&& +e^{\O_1+\O_2+\O_3} A_{12}A_{13}A_{23} +\cdots \nonumber\\
&&+e^{\O_1+\O_2+\O_3+\O_4} A_{12}A_{13}A_{14}A_{23}A_{24}A_{34}
\eea
for solitons $a$, $2n-a$, $b$, $2n-b$, with
\bea
e^{\O_1} &=&e^{\O_a}\o^{ja}   ~~~,~~~ e^{\O_2} =e^{\O_a}\o^{-ja}
\nonumber\\
e^{\O_3} &=&e^{\O_b}\o^{jb}   ~~~,~~~ e^{\O_4} =e^{\O_b}\o^{-jb}
\eea
and
\beq
A_{12} = A_{a,2n-a} ~~~,~~~A_{13} = A_{ab} ~~~,~~~A_{14}= A_{a,2n-b}~~~,
\cdots
\eeq

The classical masses of the solitons in the $c_n^{(1)}$ theory are obtained
from expressions for the energy and
momentum of the Toda field theory by substituting in these expressions the
soliton solutions \cite{Hol92}.
%The energy is given by
%\beq
%M\c (v) =\int dx \left[ \frac{1}{2} \s^2 (1+v^2) \sum_{j,k} \frac{4\vec{\a}_j
%\cdot
%\vec{\a}_k}{\a_j^2 \a_k^2} I_{jk} - \frac{m^2}{\b^2} \sum_{j=0}^n n_j\left(
%e^{i\b \vec{\a}_j \cdot \vec{\phi}} -1 \right) \right]
%\eeq
%and the momentum by
%\beq
%M\c (v) v =  \int dx \s^2v \sum_{j,k}\frac{4\vec{\a}_j \cdot
%\vec{\a}_k}{\a_j^2 \a_k^2} I_{jk}
%\eeq
%where
%\beq
%I_{jk} = \frac{ (\d_je^{\O}+2Ae^{2\O})(\d_ke^{\O}+2Ae^{2\O})}
%{(1+\d_je^{\O}+Ae^{2\O}) (1+\d_ke^{\O}+Ae^{2\O})}
%\eeq
%with $A$ given by (2.15) and $\c (v) =(1-v^2)^{-\frac{1}{2}}$. Eliminating
%$I_{jk}$ between these
%two equations gives
This leads to
\beq\label{massf}
M= -2 (1-v^2)^{-\frac{1}{2}}\,\frac{m^2}{\b^2} \int dx \sum_{j=0}^n n_j
\left( e^{i\b \vec{\a}_j \cdot \vec{\phi}} -1 \right).
\eeq
{}From the equations of motion \reff{eom} we obtain
\beq
\left( e^{i\b
\vec{\a}_j \cdot \vec{\phi}} -1 \right)= \frac{2}{m^2n_j\a_j^2}
\frac{D^2\tau_j}{\tau_j^2}
\eeq
For the $c_n^{(1)}$ theory we have $n_j\a_j^2 =2$ and
\beq
D^2\tau_j = \s^2(v^2-1)\left( e^{\O}\d_j +4e^{2\O}A+e^{3\O}\d_jA \right)
\eeq
By a change of variables $y=e^{\O}$ we can rewrite the integral in
\reff{massf} as
\bea
\int_{-\infty}^{\infty} dx \left( e^{i\b \vec{\a}_j \cdot \vec{\phi}} -1
\right)
&=& \int_0^{\infty} dy \frac{\s (v^2-1)}{m^2} \frac{\d_j +4Ay+\d_jAy^2}
{(1+\d_jy+Ay^2)^2} \nonumber\\
&=& \frac{\s (v^2-1)}{m^2}
\left[\frac{ -(2+\d_jy)}{1+\d_jy+Ay^2} \right]_0^{\infty}
\eea
We note that for soliton $n$, with $A^{(n)}=0$, the square
bracket evaluates to $1$, whereas
for the other solitons, with $A^{(a)}\neq 0$, it evaluates to $2$.
Using \reff{smass} and $\sum _{j=0}^n n_j =2n$ we find then the soliton
rest masses for the $c_n^{(1)}$ theory
\bea\label{cmass}
M_a &=& \frac{8n}{\b^2}m_a=\frac{16n}{\b^2}m \sin\frac{a\pi}{2n}  ~~~,~~~
a=1, \cdots , n-1 \nonumber\\
M_n &=& \frac{4n}{\b^2}m_n=\frac{8n}{\b^2}m
\eea
Up to an overall factor of $8n/\b^2$ these are
the masses of the elementary particles in the real coupling
$b_n^{(1)}$ affine Toda
theory. This reestablishes the, by now, well known fact that the classical
masses of the solitons of the
Toda theory based on an untwisted affine Ka\v{c}-Moody algebra $g^{(1)}$ are
proportional to the classical masses of the fundamental particles of
the Toda theory based on the dual untwisted Ka\v{c}-Moody algebra
$\check{g}^{(1)}$. We want to extend this
observation to the quantum level.

\sect{Quantum corrections to the soliton masses}

In this section we compute the first quantum corrections to the masses of
the solitons in the $c_n^{(1)}$ theory.
We follow the traditional procedure described for example
in the textbook by Rajaraman \cite{Raj82} and used in
ref.
\cite{Hol93a} for the case of $a_n^{(1)}$ solitons, where quantum corrections
are obtained from the zero-point fluctuations of the field around the classical
solutions. We split the lagrangian into kinetic and potential terms
\beq\label{lagrange}
{\cal L}[\vec{\phi}]=\half\int\left(\frac{d\vec{\phi}}{dt}\right)^2dx
-V[\vec{\phi}]
\eeq
\beq
V[\vec{\phi}]=\int dx\,\left(\half\left(\frac{d\vec{\phi}}{dx}\right)^2
-\frac{m^2}{\b^2}\sum_i n_i \left(e^{i\b\vec{\a}_i\cdot\vec{\phi}}-1
\right)\right)
\eeq
and  expand the potential $V[\vec{\phi}]$ to second order in the small
fluctuations $\vec{\eta}(x)=\vec{\phi}(x)-\vec{\phi}_s(x)$ around the
static classical soliton solution $\vec{\phi}_s(x)$
\beq\label{potential}
V[\vec{\phi}]=V[\vec{\phi}_s]+\int\,dx\,\half\eta^a(x) {\cal A}_{ab} \eta^b(x)
+{\cal O}(\eta^3),
\eeq
\beq
{\cal A}_{ab}=-\d_{ab}\frac{d^2}{dx^2}
+m^2\sum_i n_i \a_i^a \a_i^b e^{i\b\vec{\a}_i\cdot\vec{\phi}_s}.
\eeq
We look for a complete set of orthogonal eigenfunctions
$\vec{\eta}_k(x)$ of the operator ${\cal A}$
\beq
{\cal A}\,\vec{\eta}_k(x)=\nu^2_k\,\vec{\eta}_k(x)
\eeq
Note that $\cal A$ is not hermitean, but it is symmetric and so its
eigenfunctions are orthogonal without complex conjugation, i.e.,
\beq
\int dx \eta_k(x)\eta_{k'}(x)=\d_{k,k'}
\eeq
We expand
\beq
\vec{\eta}(x,t)=\vec{\phi}(x,t)-\vec{\phi}_s(x,t)=\intsumk
q_k(t)\vec{\eta}_k(x)
\eeq
to rewrite \reff{lagrange} as
\beq
{\cal L}[\vec{\phi}]={\cal L}[\vec{\phi}_s]+
\half\intsumk(\dot q_k(t))^2-
\half\intsumk\nu_k^2 (q_k(t))^2+\mbox{higher-order ~terms}.
\eeq
To quantize this, we promote the $q_k$ and $\dot{q}_k$ to canonical
pairs of operators. Their zero-point
fluctuations will give us the quantum corrections to the energy and thus
the rest mass of the soliton $\vec{\phi}_s$. Also the vacuum energy
obtains similar corrections and we will subtract them since only the
energy difference to the vacuum energy is relevant.
\beq\label{e39}
M=M_{classical}+\left(\half\hbar\intsumk\nu_k\right)_{soliton}
-\left(\half\hbar\intsuml\nu_l\right)_{vacuum}
+{\cal O}(\hbar^2).
\eeq

The quantization procedure outlined above is the standard technique for
obtaining the quantum corrections to the energy of the quantum state
arising from a static classical solution. It does however not explain why
there is a quantum state associated to the classical solution.
This is particularly unclear in light of the fact that the soliton solutions
in affine Toda theory are
complex, even though the fields $\phi$ in the Toda lagrangian \reff{lag} are
required to be real. To see that a quantum state is associated to every
classical solution, even complex ones, one has to use the path-integral
techniques described by Dashen {\em et al} in ref. \cite{Das74,Das75}.
One writes the Green's function
$tr (E-H)^{-1}$ in terms of a functional integral of the classical action
over the real field  configurations $\phi$
with periodic (in time) boundary conditions. One approximates this
path-integral
by saddle-point/steepest-descent techniques (which may necessitate excursions
into the complex $\phi$-plane). One then extracts the quantum energy
spectrum from the resulting expression and finds that a tower of
energy levels arises from each saddle point of the action,
i.e., from every solution of the classical equations of
motion.  In this point of view it is clear that, although the theory under
consideration
may be  described by real fields, complex solutions of the classical equations
contribute to the physical spectrum. The equation for the quantum masses
of the solitons obtained
from this approach is identical to \reff{e39}.

\subsection{The eigenmodes}

We begin by finding the orthogonal eigenfunctions
$\vec{\eta}_k(x)$ of
${\cal A}$  using the following trick: if one has a second
classical solution $\vec{\phi}_{2s}$ to the field equations \reff{e21}
then the difference
$\vec{\rho}(x,t)=\vec{\phi}_{2s}(x,t)-\vec{\phi}_s(x,t)$ satisfies the equation
\beq\label{ee}
\Box \vec{\rho} = -\frac{m^2}{i\b} \sum_{j=0}^n n_j\vec{\a}_j \left( e^{i\b
\vec{\a}_j \cdot
(\vec{\phi}_s+\vec{\rho} )} -e^{i\b \vec{\a}_j \cdot \vec{\phi}_s} \right)
\eeq
If the time-dependence of $\rho$ can be Fourier expanded in the
form
\beq\label{exp}
\vec{\rho}(x,t)=\sum_{l=1}^\infty e^{il\nu t}\vec{\rho}_{(l)}(x)
\eeq
then the first Fourier component $\vec{\rho}_{(1)}(x)$ satisfies
\beq
\left(-\nu^2-\frac{d^2}{dx^2}\right)\vec{\rho}_{(1)}=
-m^2\sum_i
n_i\vec{\a}_i(\vec{\a}_i\cdot\vec{\rho}_{(1)})
e^{i\b\vec{\a}_i\cdot\vec{\phi}_s}
\eeq
because ${\cal O}(\rho^2)$ terms do not contribute.
Thus $\vec{\eta}(x)=\vec{\rho}_{(1)}(x)$ is an eigenfunction of ${\cal A}$ with
eigenvalue $\nu$.

In our case we  choose $\vec{\phi}_{2s}$ to be a two-soliton solution
where one of the solitons is
described by $\vec{\phi}_s$. Thus, for any soliton $\vec{\phi}^{(a)}$ we
 consider the functions
\beq
\vec{\rho} = \vec{\phi}^{(ab)} - \vec{\phi}^{(a)}
\eeq
The complete set of eigenfunctions of $\cal A$ is obtained by considering
all values  of
$b$, and in addition letting soliton $b$ have all possible ( real and
imaginary) momenta, restricted only by the condition that $\rho$ should
be bounded.

We write the two-soliton solution as in \reff{ctwosol}, with
\bea
\O_a &=& m_a x \nonumber\\
\O_b &=& i(kx+\n t)-\xi
\eea
i.e. we put the soliton $a$ at rest at the origin, while for soliton $b$
the frequency of oscillation is, according to \reff{e27},\reff{smass},
\beq\label{e315}
\n ^2 = k^2 +m_b^2.
\eeq
Here $k$ can be real or imaginary, but $\n$ will turn out to be
real. We find then from \reff{intfunc}
\bea
A_{12} &=& \cos^2 \frac{a\pi}{2n}=A^{(a)} ~~~,~~~
A_{34}=\cos^2\frac{b\pi}{2n}=A^{(b)}
\nonumber\\
A_{13} &=&A_{24}=A_+~,~~~A_{14}=A_{23}=A_-\\
A_\pm&=&-\frac {m_a^2+m_b^2 -m_{a\mp b}^2
-2ikm_a} {m_a^2+m_b^2 -m_{a\pm b}^2 +2ikm_a}\nonumber
\eea
Defining $\d \tau = \tau^{(ab)} - \tau^{(a)}$ we expand
\beq
\vec{\rho}= \frac{i}{\b}\sum_{j=0}^n \frac{2}{\a_j^2}\vec{\a}_j \left[ \frac{\d
\tau_j}
{\tau_j} +{\cal O}\left( \left[ \frac{\d \tau_j}{\tau_j}\right]^2\right)\right]
\eeq
Here, from the two-soliton solution,
\bea
\d \tau_j &=&e^{i(kx+\n t)-\xi}\left[\d_j^{(b)}+\left(\d_j^{(a+b)}A_+
+\d_j^{(a-b)}A_-\right)e^{m_ax}
+\d_j^{(b)}A_+A_-A^{(a)}e^{2m_ax} \right] \nonumber\\
&&+e^{2i(kx+\n t)-2\xi}A^{(b)}\left[ 1+ \d_j^{(a)}A_+A_- e^{m_ax}
+A_+^2 A_-^2 A^{(a)} e^{2m_ax} \right]
\eea
and
\beq\label{tauj}
\tau_j =1+\d_j^{(a)}e^{m_ax} +A^{(a)} e^{2m_ax}
\eeq
Thus we see that $\rho$ does indeed have an expansion of the form
\reff{exp}  and we obtain the eigenfunctions of $\cal A$
\bea
\vec{\eta}_a^{(b)}&=&\vec{\rho}_{(1)}
=\frac{i}{\b} \sum_{j=0}^n \frac{2}{\a_j^2}
\vec{\a}_j\tilde{\eta_j} \\
\label{etaj}\label{e321}
\tilde{\eta}_j &=& \frac {\d_j^{(b)}
+\left(\d_j^{(a+b)}A_+ + \d_j^{(a-b)}A_-\right) e^{m_ax}
+\d_j^{(b)}A_+ A_- A^{(a)} e^{2m_a x} }
{1+ \d_j^{(a)} e^{m_a x} +
A^{(a)} e^{2m_a x} }e^{ikx-\xi}
\eea
The parameter $\xi$ determines only the normalization of the
eigenmodes.

We have obtained all eigenfunctions
of the second order differential operator $\cal A$. This follows from the
fact that we have $n$ solutions $(b=1\dots n)$ for any possible asymptotic
behaviour at $x=-\infty$
\beq
\lim_{x \rightarrow -\infty}\tilde{\eta_j}=\d_j^{(b)} \,e^{ikx-\xi}
\eeq
and the fact that the vectors
$\sum_{j=0}^n \frac{2}{\a_j^2} \d_j^{(b)} \vec{\a}_j,
{}~b=1,\dots,n$ are a complete basis of the root space.

We require that $\eta$  be bounded.
This will be true for any real values of $k$ ("scattering states" to be
treated in the next subsection) and for
some discrete imaginary values of $k$ ("bound states" to be treated
in subsection \ref{bound}) determined by zeroes of $A_\pm(k)$.

\subsection{Scattering states contributions}

The sum over contributions from real values of $k$ is done in the same
manner as in section 3.2 of ref. \cite{Hol93a} by Hollowood,
with minor modifications for the case of  the $c_n^{(1)}$ theory. We summarize
the procedure
for completeness and
refer the reader to that reference for further details.

All real values of $k$ give acceptable periodic solutions, with the
frequency determined by \reff{e315}. However, it is convenient to discretize
the sum over zero
point energies (and thus eliminate  an infrared divergence ),
by putting the soliton system in a box of size $L$ and imposing suitable
boundary
conditions on the solutions.
For this purpose it is simpler to place the center-of-mass of soliton
$\phi^{(a)}$ at $x=L/2$, thus shifting in \reff{etaj} $m_ax \rightarrow
m_a (x-
\frac{L}{2})$, and impose boundary conditions that the solutions vanish at
$x=0$ and $x=L$.

Acceptable solutions that vanish at $x=0$ are given by
\beq
\tilde{\eta} (k,x) -\tilde{\eta}(-k,x)
\eeq
Requiring that the solutions vanish also at $x=L$  (in the limit of large
$L$) restricts the values of $k$ according to
\beq\label{e324}
k_pL+ \rho_b(k_p) =\pi p ~~~,~~~ p \in Z
\eeq
where
\beq
\rho_b (k)= -\frac{i}{4}\e_a \ln \frac{A_+(k) A_-(k)}{A_+(-k) A_-(-k)}
\eeq
where $\e_a=2-\d_{a,n}$.

As discussed in ref. \cite{Hol93a} the sum over the modes diverges
and two renormalizations are required. The first one, as written in \reff{e39},
 corresponds to
subtracting from the
zero-point energies of the modes around the soliton solution the zero point
energies of the modes around the vacuum (corresponding to $\rho_b (k)=0$ in
\reff{e324}). This removes a quadratic divergence. Thus, the scattering states
contributions to the soliton mass are given by the sum
\bea
\D M_a &=& \frac{1}{2} \sum_{b=1}^n \sum_{k_p \geq 0} \left[ \sqrt{
k_p^2+m_b^2} - \sqrt{ [k_p+\r _b(k_p)/L]^2 +m_b^2} \right]
\nonumber\\
&=&-L \sum_{b=1}^n \int_{-\infty}^{\infty} \frac{dk}{4\pi} \frac{\r_b(k)}{L}
\frac
{d\e_b(k)}{dk} \nonumber\\
&=& -\frac{1}{4\pi} \sum_{b=1}^n \left[ \r_b(k)\e_b(k) |_{-\infty}^{\infty}
+\int_{-\infty}^{\infty} dk \e_b(k) \frac{d\r_b(k)}{dk} \right]
\eea
Following Hollowood we have taken $L \rightarrow \infty$ and replaced the
sum over $p$ by an integral over $k$
\beq
\sum_{p \geq 0} = L \left[ \int_{-\infty}^{\infty} \frac{dk}{2\pi}
 +O(L^{-1})\right]
\eeq
with $\e_b(k) = \sqrt{k^2+m_b^2}$.

A second renormalization is required in order to remove a logarithmic
divergence in the
integral.  As discussed by Hollowood it is achieved by normal-ordering the
original Toda
Hamiltonian.
We arrive at the following expression for the corrections from scattering
states to the mass of soliton $a$ (corresponding to eq. (3.25) in
\cite{Hol93a})
\beq\label{e328}
\D^{scattering} M_a=m_a\e_a\left(-\frac{n+1}{2\pi}+\sum_{b=1}^n\O_b\right)
\eeq
where
\bea
\O_b &=&
 \int_{-\infty}^{\infty} \frac{dk}{4\pi}\left(
\half\left(\frac{m_b}{m}\right)^2\frac{1}{\sqrt{k^2+m_b^2}} \right.\\
 &-& \left. \sqrt{k^2+m_b^2} \left[\frac{2(m_a^2+m_b^2-m_{a+b}^2)}
{(m_a^2+m_b^2-m_{a+b}^2)^2+4m_a^2 k^2} +
\frac{2(m_a^2+m_b^2-m_{a-b}^2)}
{(m_a^2+m_b^2-m_{a-b}^2)^2+4m_a^2 k^2}\right]
\right) \nonumber
\eea
The first term in the integrand is the normal-ordering counterterm.

The integral is straightforward to perform. Substituting the explicit
expressions in (2.8) for the masses and making the change of variable
\beq
k= m_b \frac{u}{\sqrt{1-u^2}}
\eeq
we obtain
\beq
\Omega_b=\frac{1}{\pi} \frac{\sin \frac{b\pi }{2n}}{\sin \frac{a\pi }{2n}}
\frac{\cos \frac{ (a+b)\pi}
{2n} }{| \cot\frac{  (a+b)\pi}{2n} |}
\tan ^{-1} \frac{1}{| \cot \frac{ (a+b)\pi}{2n} |}  + \Omega_b^{\infty}
\eeq
where the  inverse tangent  is evaluated on the principal branch between $0$
and $\pi /2$ and
\beq
\Omega_b^{\infty}=\frac{1}{2\pi}\cot \frac{a\pi}{n} \sin \frac{2b\pi}{n}
 \int_{-1}^{1}\frac{du}{1-u^2}
\eeq

In the sum of   \reff{e328}  the terms coming from $\Omega_b^{\infty}$ cancel,
and a
tedious calculation yields then
\beq\label{scmass}
\D^{scattering}
M_a=m_a\e_a\left(-\frac{n+1}{2\pi}-\frac{1}{8}\cot\frac{\pi}{2n}
+\half\cot\frac{a\pi}{2n}\left(\frac{a}{2n}-\half\right)\right).
\eeq

\subsection{Bound states contributions}\label{bound}

Additional contributions to the mass corrections come from imaginary values of
$k$ which give bounded modes. We consider first the case $a \neq n$.
For imaginary values of $k$ the asymptotic values of the $\tilde{\eta}_j$
of \reff{etaj}
can remain finite at both $x= \pm \infty$ if either $A_+(k)=0$ or
$A_-(k)=0$. Writing
\beq
A_\pm(k) = - \frac{\a_\pm -ik}{\a_\mp +ik}
\eeq
with
\beq
\a_{\pm} =\frac{m_a^2+m_b^2-m_{a \mp b}^2}{2m_a} = \pm m_b \cos \frac{
\pi (a \mp b)}{2n}
\eeq
 we must have $k=-i\a_{\pm}$. At zeroes of
$A_+(k)$ the asymptotic forms of the modes are
\bea
\lim_{x \rightarrow \infty} \tilde{\eta}_j (k= -i\a_+) &=&
e^{(m_b \cos\frac{\pi (a-b)}{2n} -m_a)x} \d_j^{(a-b)}
\frac{A_-}{A^{(a)}}\\
\lim_{x \rightarrow - \infty} \tilde{\eta}_j (k= -i\a_+)&=&
e^{m_b \cos \frac{\pi (a-b)}{2n}x} \d_j^{(b)}\label{as}
\eea
Thus, the modes will be bounded if
\bea
(i)~~~&& \sin \frac{\pi b}{2n} \cos \frac{\pi (a-b)}{2n}
-\sin \frac{\pi a}{2n} \leq 0 \\
(ii)~~~&& \sin \frac{\pi b}{2n} \cos \frac{\pi (a-b)}{2n} \geq 0
\eea
i.e. if
\beq\label{e337}
b \leq a  ~~~~{\rm or} ~~~~ b=n
\eeq
Actually the case $b=a$ corresponds to the translational zero mode
and should not be included here. The case $b=n$ is special because
$A_+=A_-=0$ so that its asymptotic behaviour at $x\rightarrow\infty$
is given by
\beq
\lim_{x \rightarrow \infty} \tilde{\eta}_j (k= -i\a_+) =
e^{(m_b \cos\frac{\pi (a-b)}{2n} -2m_a)x} \frac{\d_j^{(b)}}{A^{(a)}}
\eeq
rather than \reff{as}.

In a similar manner, at zeros of $A_- (k)$ we have
\bea\label{prob}
\lim_{x \rightarrow \infty} \tilde{\eta}_j (k= -i\a_-)&=&
e^{(-m_b \cos \frac{\pi (a+b)}{2n}-m_a)x} \d_j^{(a+b)}
\frac{A_+}{A^{(a)}}\\
\lim_{x \rightarrow - \infty} \tilde{\eta}_j (k= -i\a_-)&=&
e^{-m_b \cos\frac{\pi (a+b)}{2n}x}\d_j^{(b)}
\eea
giving the conditions
\bea
(i)~~~&& \sin \frac{\pi b}{2n} \cos \frac{\pi (a+b)}{2n}
+ \sin \frac{\pi a}{2n} \geq 0 \\
(ii)~~~&& \sin \frac{\pi b}{2n} \cos \frac{\pi (a+b)}{2n} \leq 0
\eea
i.e.
\beq\label{e340}
b \geq n-a
\eeq
The case $b=n$ leads to the same state as the $b=n$ case in \reff{e337} and
we must count it only once. One might expect that
further bounded solutions may arise from \reff{e321} at poles of $A_\pm$ by
choosing $\exp{\xi}=A_\pm$. Repeating the preceding analysis one finds
however that the state obtained from a pole of $A_\pm$ in the
perturbation by soliton $b$ is equal to the state obtained from
a zero of $A_\pm$ in the perturbation by soliton $n-b$.
Thus we obtain
\beq\label{e347}
\D^{bound} M_a=\half\sum_{b=1}^{a-1} \nu_+^{(b)}+
\half\sum_{b=n-a}^{n}
\nu_-^{(b)}~,~~~a=1,\dots,n-1
\eeq
where the frequencies are obtained from \reff{e315}
\beq
\nu_\pm^{(b)}=2m\sin\frac{b\pi}{2n}\left|\sin\frac{(a\mp b)\pi}{2n}\right|
\eeq
Performing the sum gives the result
\beq\label{bmass}
\D^{bound} M_a= \half m_a\left(\cot\frac{\pi}{2n}
+\cot\frac{a\pi}{2n}\right)~,
{}~~~a=1,\dots,n-1
\eeq

The case $a=n$ has to be treated separately because $A^{(n)}=0$. By
similar considerations as the above one finds that
\beq\label{e344}
\D^{bound} M_n=\sum_{b=1}^n\nu_+^{(b)}=
\frac{1}{4} m_n\cot\frac{\pi}{2n}.
\eeq

\subsection{Result}

Combining  the classical mass in  \reff{cmass}, the corrections from the
scattering states  in \reff{scmass} and those from the bound states  in
\reff{bmass},\reff{e344}
we obtain the quantum soliton mass to first order
\beq\label{result}
M_a^{soliton}=\e_a  m \sin\frac{a\pi}{2n}
\left(\frac{8n}{\b^2}-\frac{n+1}{\pi}
+\frac{1}{4}\cot\frac{\pi}{2n}+\frac{a}{2n}\cot\frac{a\pi}{2n}\right)
+{\cal O}(\b^2),
\eeq
where $\e_a=2-\d_{a,n}$.
Thus the mass ratios receive the following quantum corrections
\beq\label{ratios}
\D\left(\frac{M_a}{M_b}\right)=
\frac{m_a}{m_b}\frac{\e_a}{\e_b}\frac{\b^2}{16n^2}
\left(a\cot\frac{a\pi}{2n}-b\cot\frac{b\pi}{2n}\right)
+{\cal O}(\b^4).
\eeq

\sect{Discussion}

Classically it has been known  for some time that the soliton masses in the
 $c_n^{(1)}$ theories
are proportional to the masses of the fundamental particles of the $b_n^{(1)}$
theories. We can now check this soliton -- particle duality at the quantum
level.

For the  $c_n^{(1)}$ Toda theory with {\em imaginary} exponentials
\beq
{\cal L}=-\half \vec{\phi} \cdot \Box \vec{\phi} + \frac{m^2}{\b^2}
 \sum_i n_i\,\left(e^{i\b \vec{\a}_i ^{~}\cdot\vec{\phi}}-1\right)
\eeq
we have found the mass corrections, which we rewrite to the order of our
approximation as
\beq
M_a^{soliton}=\e_a  m       \left(\frac{8n}{\b^2}-\frac{n+1}{\pi}
+\frac{1}{4}\cot\frac{\pi}{2n}\right)   \sin\frac{a\pi}{2n}
\left(1   +\frac{a}{2n}\cot\frac{a\pi}{2n}\right)
+{\cal O}(\b^2),
\eeq

On the other hand, for the $b_n^{(1)}$ Toda theory with {\em real} exponentials
\beq
{\cal L}=-\half \vec{\phi} \cdot \Box \vec{\phi} - \frac{\mt^2}{\b^2}
 \sum_i \check{n}_i\,\left(e^{\b \check{\vec{\a}}_i \cdot\vec{\phi}}-1\right)
\eeq
the full quantum masses of the fundamental particles
have been found in \cite{Del92b} to be
\bea
M_a^{particle}(\bn,\mt,\b^2)&=&\e_a
\mt\sin\left(\frac{a\pi}{2n-\frac{\b^2/4\pi}{1+\b^2/4\pi}}
\right) \nonumber\\
&=& \e_a \mt\sin\frac{a\pi}{2n}\left(1
+\frac{a}{2n}\cot\frac{a\pi}{2n}\right)
+{\cal O}(\b^2),
\eea
Comparing this with our result  for the soliton masses in the
$\cn$ theories we find the relationship
\beq\label{massrel}
M_a^{particle}(\bn,\mt,\b^2)=M_a^{soliton}(\cn,m,\b^2)+{ \cal O}(\b^2)
\eeq
provided the mass scales are related by
\beq
\mt=m\left(\frac{8n}{\b^2}-\frac{n+1}{\pi}+
\frac{1}{4}\cot\frac{\pi}{2n}\right).
\eeq
In particular for the mass ratios
\beq
\frac{M_a^{particle}}{M_b^{particle}}(\bn,\b^2)=
\frac{M_a^{soliton}}{M_b^{soliton}}(\cn,\b^2)+{ \cal O}(\b^4).
\eeq
Thus the mass ratios of the solitons of the
imaginary coupling
$\cn$ Toda theories are equal to the mass ratios of the fundamental particles
of the
real coupling
$\bn$ Toda theory, even after taking the first quantum corrections into
account. We  expect this relation to hold at the full quantum level, to all
order in $\b^2$.

If we specialize our result to the case of $c_2^{(1)}$ we can
compare to the calculation of Watts \cite{Wat94}. It would appear that
for the $n=2$ theory Watts omitted, in the mass
corrections for soliton $a=1$, the contribution from the
$b=n-a=1$ state in the second sum of \reff{e347}. It is an accident of
the $n=2$ case that omission of this contribution still leads
to reasonable results (except that the masses of the solitons turn
out to be proportional to those of the fundamental particles
in the  $b_2^{(1)}$ theory with {\em imaginary} coupling rather
than the theory with {\em real} coupling). However, in the general
$n$ case such an omission would lead to  completely unreasonable
results for the soliton mass corrections, which would not be
related to the masses of fundamental particles of any theory.

The equality between the soliton masses $M_a^{soliton}(\cn,m,\b^2)$ and the
particle masses  $M_a^{particle}(\bn,\mt,\b^2)$ does not imply that the
solitons of the $\cn$ theory can be reinterpreted as the fundamental particles
of the $\bn$ theory. One  feature  that stands in the way of such an
interpretation is the fact that the solitons have a multiplet structure (i.e.
there are several solitons of the same mass) whereas the fundamental particles
do not. The corresponding problem  was noted already in the original paper
\cite{God77} on the monopole -- gauge particle duality. In that case the gauge
particles have a multiplet structure but the monopoles do not. In sine-Gordon
theory ($a_1^{(1)}$ Toda theory) it is however known that a breather state
(soliton -- antisoliton bound state) can be identified with the fundamental
particle of that theory. This would make it interesting to extend our
calculations to the
breathers of Toda theory.

Using the strong coupling -- weak coupling duality discovered in \cite{Del92b}
we can extend the relationship \reff{massrel} to the following squares of
relations
\beq\label{relations}
\begin{array}{ccc}
\frac{M_a^{particle}}{M_b^{particle}}(\bn,\b^2)&=&
\frac{M_a^{soliton}}{M_b^{soliton}}(\cn,\b^2)\\
\|&&\|\\
\frac{M_a^{particle}}{M_b^{particle}}(\an,(4\pi/\b)^2)&=&
\frac{M_a^{soliton}}{M_b^{soliton}}(?\an,(4\pi/\b)^2)\\
\\
\frac{M_a^{particle}}{M_b^{particle}}(\cn,\b^2)&=&
\frac{M_a^{soliton}}{M_b^{soliton}}(?\bn,\b^2)\\
\|&&\|\\
\frac{M_a^{particle}}{M_b^{particle}}(\dn,(4\pi/\b)^2)&=&
\frac{M_a^{soliton}}{M_b^{soliton}}(?\dn,(4\pi/\b)^2)
\end{array}
\eeq
The algebras which are preceeded by a question mark in the above relations have
been identified by classical calculation only. It would clearly be desirable to
have quantum calculations for the solitons of all affine algebras to confirm
this diagram  and similar ones involving the remaining Toda theories.

%\vspace{1cm}

Looking at the relations \reff{relations} one wonders what really is the full
algebra which underlies the $\cn$ Toda theory at the quantum level. It seems to
 be some $\b$-dependent amalgamate of $\cn$, $\bn$, $\an$ and $\dn$. It would
be
very interesting to unearth this structure from the quantization of affine
Toda theory.

An interesting part of this algebraic structure has been described by Bernard
and LeClair \cite{Ber91}. Applying their formalism one observes that the $\cn$
affine Toda theory has a $U_q(\dn)$ symmetry. Here $U_q(\dn)$ is the quantum
deformation \cite{Dri86} of the enveloping algebra of the affine Kac-Moody
algebra $\dn$. The deformation parameter $q$ is related to the coupling
constant by $q=\exp{(8i\pi^2/\b^2)}$. The quantum solitons are expected to
transform in
some finite dimensional representation under this quantum affine algebra. This
implies that the soliton scattering matrices, which also have to respect this
symmetry, are proportional to the corresponding R-matrices of the quantum
affine algebra. The soliton S-matrices for $a_n^{(1)}$ Toda theory have been
obtained this way \cite{Hol93b}.

However, this can not be the whole story. In the present work we have shown
that the
mass ratios of the $\cn$ quantum solitons receive quantum corrections. But it
is
known that any S-matrices obtained from R-matrices for the finite dimensional
representations of quantum affine algebras predict  $\b$-independent
(classical)
mass ratios; these ratios are determined by  pole locations,
and the location of  poles of  R-matrices is fixed by
the values of the Casimir numbers of the algebra in the finite dimensional
representations \cite{Del94a}.

Recently R-matrices with moving poles have been discovered \cite{Bra94}. They
arise from an enlargement of quantum affine algebras, namely from type I
quantum affine superalgebras \cite{Del94b}. This suggests that the problem of
finding the S-matrices for the solitons with $\b$-dependent mass ratios might
find its solution through the identification of some, as yet unknown, larger
quantum symmetry algebra. This algebra should at the same time explain the
relations \reff{relations}.

While finding the soliton S-matrices for non-selfdual algebras is still an open
problem, the
corresponding problem for the fundamental particles has been solved. In all
affine Toda theories based on non-selfdual Kac-Moody algebras, the mass ratios
of the fundamental particles renormalize. Therefore the S-matrix formulas
obtained in \cite{Bra90}, which have been given an elegant Lie algebraic
description in \cite{Dor92}, do not apply. The correct formulas
\cite{Del92b,Cor93}
contain a $\b$-dependent Coxeter number interpolating between the integer
Coxeter numbers of dual Kac-Moody algebras. In our case it interpolates between
$\cn$ and $\dn$. This should be part of the algebraic structure to be
discovered.

%One interesting aspect of the solitons in affine Toda theory is that they are
%complex, even though the fields $\phi$ in the Toda lagrangian \reff{lag} are
%required to be real. There has been some debate as to what the relevance of
%complex solutions for real fields might be. One can not simply declare the
%lagrangian \reff{lag} to be a lagrangian for complex fields because the
%corresponding action would not be bounded from below. Our view on this matter
%is the following:

%Toda theory is a theory of real fields and it is quantized by performing a
%%path
%integral over real field configurations. However, to evaluate this real
%integral one can resort to the path-integral analogue of a method familiar
%%from
%functional analysis: one can deform the integration contour into the complex
%plane. In particular, if we want to approximate the path integral by
%%stationary
%phase approximation, then we have to choose the integration contour to pass
%through all the saddle points in the complex plane. Each saddle point gives a
%contribution to the path integral and thereby eventually to the spectrum of
%%the
%quantum theory. The complex soliton solutions are these saddle points. The
%quantum corrections which we have calculated in this paper arise from the
%harmonic approximation of the integral near these saddle points. Having seen
%the relevance of complex solutions in Toda theory, it might be profitable to
%look out for this possibility in other physical theories as well.

There are clearly many lessons to be learned from the quantum theory
of affine Toda theory. We have two goals in mind in pursuing these studies: we
hope to learn from these integrable theories how to handle soliton effects in
other physical theories and we hope to discover the new algebraic structure
describing the quantum symmetry of these theories.

We thank Mike Freeman for discussions.

{\bf Note added:} Watts has explained to us why he omits the $b=n-a=1$ state
in his calculation \cite{Wat94} of the $c_2^{(1)}$ soliton masses:
At exactly the location at which $A_-$ has the zero which leads to this
bound state, $A_+$ has a pole, leading to a problem in \reff{prob}. We
argue that this coincidence of the zero of $A_+$ with a pole of $A_-$ is
an artifact of the classical approximation and does not hold in the
quantum theory. In the quantum theory the poles and zeros of the
interaction functions $A_{ab}$ become the poles and zeros of the
soliton a -- soliton b S-matrix. Their location is determined by the
quantum soliton and breather mass ratios. As we have seen in this paper,
these soliton mass ratios receive quantum corrections and thus the
poles and zeros move. It will be interesting to analyze this in detail.

MacKay and Watts have published independently from us on the
same day a paper \cite{Mac94} calculating the first mass corrections for the
solitons of all affine Toda theories. For the non-selfdual Toda theories they
find that the soltion mass ratios do {\it not} renormalize like the
fundamental particles of the dual theories. They are thus in contradiction
to our result in this paper for the $c_n^{(1)}$ Toda theory. We belive
that this discrepancy may be due to the same b=n-a bound state
mentioned above.

\end{document}